# Interference Allocation Scheduler for Green Multimedia Delivery

Siyi Wang, Weisi Guo[†], Chadi Khirallah[*], Dejan Vukobratović[‡], and John Thompson[*] *Member, IEEE*

**Abstract**

One of the key challenges in wireless networking is how to allocate the available radio resources in order to maximise key service delivery parameters such as the aggregate throughput and the multimedia quality of experience (QoE). We propose a novel and effective scheduling policy that allocates resource blocks, such that interference power is shifted towards capacity-saturated users, while improving the throughput of unsaturated users. The highlight of the research is that the proposed scheme can dramatically improve the performance of cells that have a high discrepancy in its signal-to-noise-ratio (SNR) distribution, which is typical in urban areas. The results show that a *free-lunch* solution is possible, whereby for a negligible performance degradation in the saturated users, a large improvement in the non-saturated users can be obtained. However, on average, the number of free-lunch user pairings are low. By relaxing the degradation constraints, the *non-free-lunch* solution can yield a greater multi-user throughput gain. Motivated by surge in mobile multimedia traffic, we further demonstrate that the proposed scheduling may have a profound impact on both energy efficiency and QoE of multimedia service delivery.

## I. INTRODUCTION

### A. Motivation: Green Multimedia Delivery

Over the past decade, mobile traffic has been transformed from mainly voice-based to a heterogeneous amalgamation of different data types. As of 2012–2013, the main driver of mobile content is multimedia data. This is recently confirmed by the Cisco Visual Networking Index (VNI), which estimates that mobile video traffic will increase 16-fold in the period 2012–2017, accounting for almost 67% of all mobile data traffic by 2017 [1]. As a result of this traffic growth, operators are increasingly expending energy to transmit more information across greater coverage areas. The resulting operational expenditure (OPEX) and carbon footprint of the growth is proving to be costly [2]. It is estimated that 15% of a cellular operator's OPEX is attributed to the energy consumption of its base stations (BSs) [3], [4]. Therefore, there is a genuine commitment from industry to reduce the energy expenditure and the resulting carbon footprint [5]. Another dominating trend is the fact that for the first time in history, more people live in cities. Therefore, the urban mobile data demand is set to rise rapidly due to both the desire for greater digital connectivity and the increasing population density in most cities around the world.

Siyi Wang is with Institute for Telecommunications Research, University of South Australia, Australia, (e-mail: siyi.wang@mymail.unisa.edu.au)
Weisi Guo[†] is with School of Engineering, The University of Warwick, UK, (e-mail: weisi.guo@warwick.ac.uk)
Chadi Khirallah[*] and John Thompson[*] are with School of Engineering, The University of Edinburgh, UK, (e-mail: {c.khirallah, john.thompson}@ed.ac.uk)
Dejan Vukobratović[‡] is with Department of of Power, Electronics and Communication Engineering, University of Novi Sad, Serbia, (e-mail: dejanv@uns.ac.rs)

## B. Review: Energy- and Spectral-Efficiency

Several academic papers have been performed on the network level to show that energy- and spectral- efficiency improvements are possible. A number of studies have characterized the trade-off between energy- and spectral-efficiency for single links and for video streaming in mobile user scenarios [6]. Other research has focused on adjusting the operation of the network with respect to dynamic traffic patterns [7], [8]. Combining the growth of multimedia content and the pressure to reduce energy consumption, there is a recognised need to improve multimedia content delivery [6], [9], [10] and to do so in an energy efficient manner [11].

## C. Review: Interference Management

Interference management is a key issue in interference-limited cellular networks, such as those found in dense urban areas. This area has been tackled from a variety of research angles, and can be categorised into the following areas [12]: i) cancellation, ii) reduction, and iii) avoidance. We will primarily focus on the scheduling techniques used to avoid interference between co-frequency transmitters. It has been demonstrated that interference avoidance can be partially and completely achieved through coordinated scheduling techniques between transmitter pairs on a dynamic level using either pre-set or learning algorithms [13], [14]. This has a similar philosophy and performance to hard- or soft-frequency-reuse (SFR) schemes [15]. The caveat with interference avoidance is that it incurs a significant throughput penalty at high traffic loads [13], which is reasonable given that most cells incur a low traffic load (more than half of the data traffic is carried by less than 10% of the cells).

The potential benefits of the scheduling approach to interference management are numerous. Interference cancellation schemes typically require multiple antennas to be highly effective, with most studies citing 4–8 MIMO transceivers on each BS [16], which is an expensive investment. Furthermore, interference cancellation benefits only relate to the cell-edge region, where the average received signal power from two or more cells is roughly similar. Scheduling schemes on the other hand improve the performance of most of the cell's coverage area and require minimal hardware investments.

The majority of existing research on scheduling schemes employs the Shannon capacity utility for throughput, which assumes no mutual information saturation. In a realistic transmission system, the discrete modulation-and-coding schemes' (MCSs) data rates saturate at certain SNR thresholds. As shown in Fig. 1, the difference between the Shannon capacity and a realistic 4G LTE throughput produced from MCSs is significant. For any parallel channel optimisation, it has been shown that the saturation effect can significantly affect the solution [17], [18]. To the best of our knowledge, existing scheduling research does not explicitly exploit the relationship between the throughput saturation and interference power.

## D. Contribution

In a typical network, 90% of the traffic load is carried by less than 50% of the cells. That is to say, there is a high degree of flexibility and slack in the RRBs in most cells. We propose to exploit the flexibility by re-assigning interfering RRBs from unsaturated to saturated channels. The objective of this paper is to improve the throughput of users (UEs) with unsaturated capacity channels (Point 2 in Fig. 1) by reducing the amount of interference experienced. The constraint is that the interference will be allocated to a saturated UE, who may or may not experience a degradation in throughput (Point 1 in Fig. 1) [18]. In the first part of the paper, we give the problem formulation, derive the expected throughput rate gain and resource allocation strategy for both a *free-lunch* (no performance sacrifice) and a *non-free-lunch* (some performance sacrifice) solution. In the second part of the paper, we focus on how UEs can be paired and how to coordinate inter-cell cooperation. In the third part of the paper, we compare the performance with other schedulers and examine their impact on multimedia service delivery QoE and energy consumption. The theoretical results are reinforced by the simulation results from an industrially bench-marked multi-cell and multi-UE dynamic simulator.

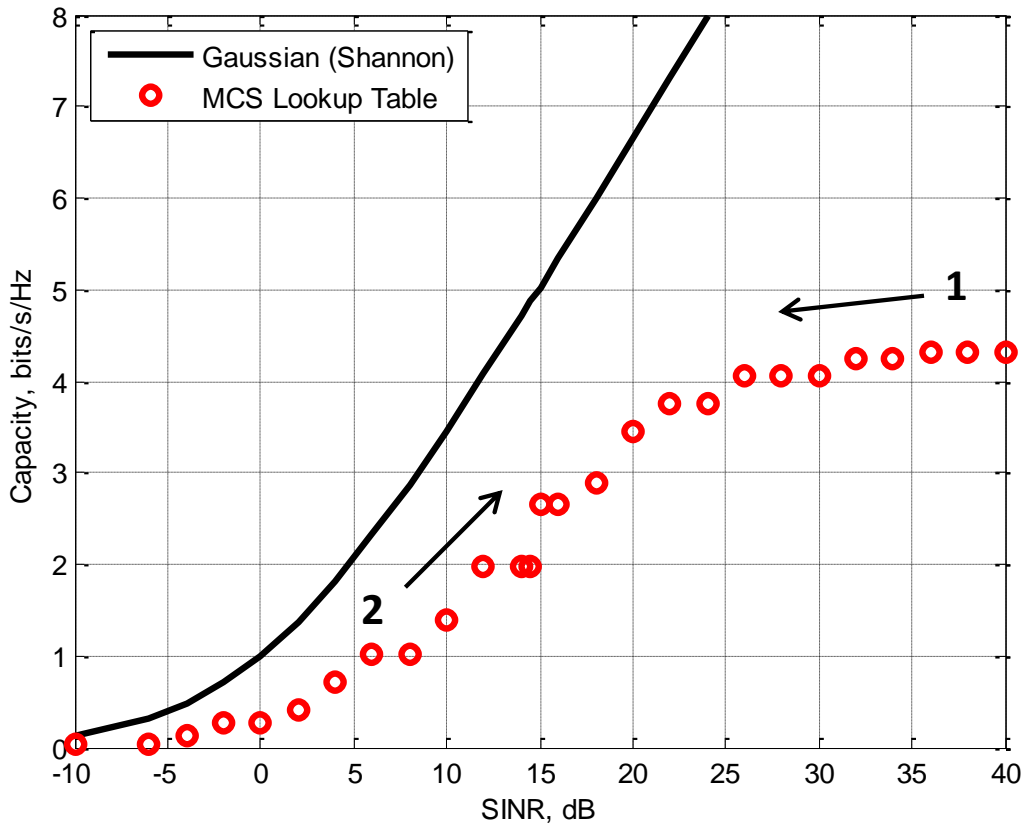

Fig. 1. Comparison from [18] showing the difference between Shannon theoretical capacity bound and simulated MCSs throughput of LTE physical layer in a 3GPP outdoor fading channel [19]. Point 1 shows a saturated capacity channel and Point 2 shows an unsaturated capacity channel.

## II. SYSTEM SETUP

### A. Multiple Base Station Network

This paper considers the downlink (DL) channels of a 4G LTE network, which utilises an orthogonal-frequency-division-multiple-access (OFDMA) system. Individual radio-resource-blocks (RRBs) can be assigned to different UEs by the scheduler. In such a multi-cell environment, co-channel RRBs will act as mutual interferers to each other. Typically in urban environments, the co-frequency BS [1] density is large ($\sim$1–2 per km$^2$), and the performance is interference-limited, as opposed to noise- or

---
[1]Each BS can have several cells or cell sectors, but this paper only considers BSs with 1 cell.

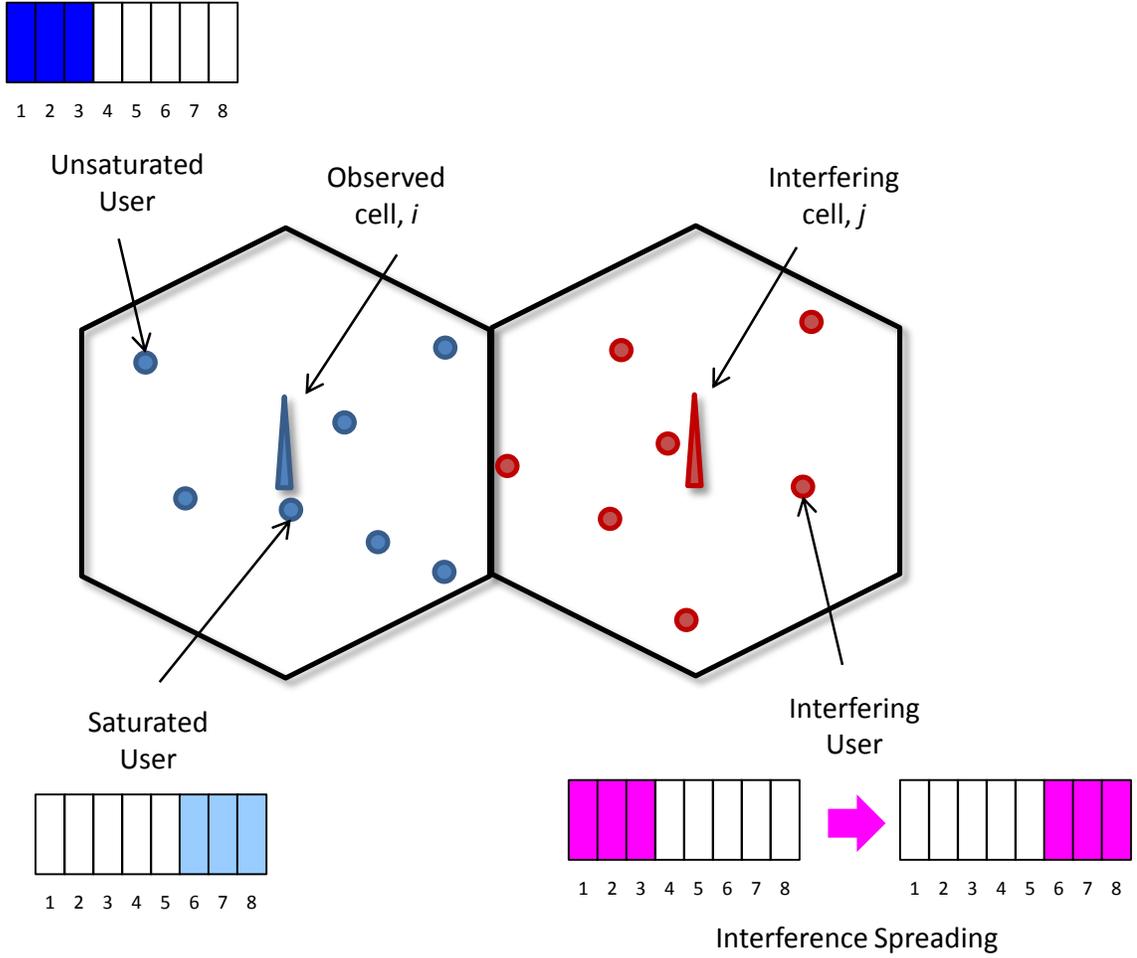

Fig. 2. Illustration of an observed serving BS and an interference BS (the paper considers many such interference BSs). In the interference BS, the scheduler can choose to reallocate the effective interference from certain RRBs to others.

propagation- limited. Therefore, a key area of research is in how to mitigate and reduce the effect of interference in order to improve the Quality-of-Service (QoS) and the Quality-of-Experience (QoE).

The cellular network consists of $(N+1)$ BSs such that the number of BSs is sufficiently large to fully account for the effects of mutual interference [20]. In order to demonstrate the mechanism and theoretical allocation bounds, two particular UEs in the observed BS shown in Fig. 2 are analysed. One UE experiences a strong channel quality and receives a saturated capacity transmission (Point 1 in Fig. 1). The other UE receives a lower channel quality and an unsaturated capacity transmission (Point 2 in Fig. 1).

We define $\gamma$ as the Signal-to-Interference Ratio (SIR) of a DL channel [2], and $C(\gamma)$ as the achievable capacity as a result of the SIR. The SIR of a UE $m$, attached to BS $i$, with $N$ interference RRBs is defined as:

$$\gamma_{m,i,N} = \frac{P_i \lambda d_{m,i}^{-\alpha} |h_{m,i}|^2 S_{m,i}}{\sum_{j=1, j \neq i}^{N+1} P_j \lambda d_{m,j}^{-\alpha} |h_{m,j}|^2 S_{m,i}}, \quad (1)$$

[2]The paper examines an interference-limited cellular network, whereby the aggregate interference power received is always far greater than the additive noise power. Therefore, in such a system the signal-to-interference plus noise ratio (SINR) is approximated by the SIR.

TABLE I
SYSTEM PARAMETERS

| Parameter | Symbol and Value |
|---|---|
| Cellular Network | OFDMA based LTE |
| Number of BSs | $(N+1)$, 19 |
| Interference Modelling | Wrap around: $\mathcal{T}$ tiers |
| BS Density | $\Lambda$, 1.15 per km$^2$ |
| BS and UE Distribution | Uniform random |
| Avg. Inter-BS Distance | $\delta$, 1 km |
| Bandwidth | $B$, 20 MHz |
| SIR | $\gamma$ |
| Average Interference Power | $I$ |
| Single Link Capacity | $C$ |
| Modulation and Coding Scheme | LTE adaptive MCS |
| Statistical Pathloss Model | 3GPP Urban Micro |
| Pathloss Distance Exponent | $\alpha$ |
| Pathloss Constant | $\lambda$ |
| Average Distance Loss | $\mathcal{D}$ |
| Multipath Fading Gain | $h$, Rayleigh |
| Shadow Fading | $S$, Log-normal |
| Shadow Fading Variance | $\sigma^2$, 9 dB |
| Traffic Load on BS | $C_{\text{traffic}}$ |
| BS Cluster Size | $(N^*+1)$ |
| Degradation Tolerance Factor | $F$ |
| Interference RRB Allocation | $X$ |
| Saturated User Capacity | $C_{\gamma_s}$ |
| Unsaturated User Capacity | $C_{\gamma_u}$ |
| Multimedia Transmitted | H.264 SVC |
| Minimum Quantisation Step-Size | $q_{\min}$, 16 |
| Maximum Frame-Rate | $t_{\max}$, 30 fps |
| Maximum Data-Rate | $R_{\max}$, 806 kbps |
| BS Transmit Power | $P$, 40 W |
| Radiohead Efficiency | $\mu$, 0.3 |
| BS Overhead Power | $P_{\text{OH}}$, 250 W |
| Energy Consumption | $E$ |

where $\lambda$ is the frequency dependent pathloss coefficient and $\alpha$ is the distance dependent pathloss exponent. The antenna gain is assumed to be omni-directional in both the azimuth and elevation planes, and thus will not be considered. The parameter $|h|^2$ is the multipath fading gain ($\sim \mathcal{X}^2$) and $S$ is the log-normal shadow fading gain with variance $\sigma^2$. Furthermore, the paper defines the following throughput terminology:

- User Capacity: the throughput achieved by a single DL transmission link of a UE;
- User-Pair Sum-Rate: the aggregate throughput achieved by a UE pairing specific to the proposed scheme;
- BS Throughput: the aggregate rate achieved by all UEs in a BS.

A full list of simulation setup parameters and symbols used in equations is shown in Table I.

## B. Review of Throughput Metrics

Existing analysis in wireless communications has largely assumed that the transmission scheme can match the Shannon bound [21]. That is to say, the capacity scales indefinitely and logarithmically with the channel quality (SNR). As shown in our earlier work [18], the performance of the adaptive MCS of the LTE physical layer performs significantly lower than the Shannon bound (Fig. 1). Furthermore, it can be seen that there is a region where the capacity is saturated (Point 1: SNR > 25 dB) and improving the channel quality leads to a negligible increase in the achieved capacity. This is due to the mutual information saturation of discrete modulation schemes [22]. Capacity saturation can occur near the serving BS location, despite being interference-limited.

There are numerous theoretical and empirical capacity expressions that improve upon the Shannon expression by considering the effects of discrete modulation schemes, coding block length, and other inefficiencies. The main drawback is that these existing expressions do not satisfy one or more of the following issues:

1) a single expression describes the entire set of MCSs of a communications system. Typically the theoretical expressions only describe a single MCS (i.e., QPSK modulation with a rate 1/3 Turbo code). This is the case for work in [22]–[24].
2) a tractable function that can be used in more complex optimisation frameworks, which is the case for work in [22], [23], [25].

Therefore, there was a need to employ a tractable empirical expression that can describe the capacity profile of adaptive MCSs shown in Fig. 1. Our earlier work [18] showed that a tractable empirical expression can be found to describe the capacity of arbitrary modulation inputs with FEC coding. We utilise this expression in this paper:

$$C(\gamma) \approx a \arctan\left(\frac{\gamma + b}{c}\right). \tag{2}$$

The SNR ($\gamma$) is a linear ratio, given by $\gamma = \frac{|h|^2 P}{N_0}$. The combined *adaptive* MCS for LTE in a 3GPP channel has a unique set of adjustment parameters: $a = 2.27$, $b = 13$, and $c = 40$. Proof of the arctangent empirical relationship backed by simulation data can be found in Appendix A and is also given in more detailed form in [18].

The rationale for choosing the inverse tangent function is that it is both tractable and it matches the capacity characteristics of the whole set of adaptive MCSs. The approximation accuracy is very high: a mean error of $2.5 \times 10^{-2}$ bit/s/Hz and a variance of $2.8 \times 10^{-3}$, across all SNR ranges.

## III. Scheduler Mechanism

As previously mentioned, the network consists of $(N+1)$ BSs. The paper divides the BSs into clusters, each with $(N^*+1)$ BSs. The interference allocation (IA) scheduler has the following key properties:

- At any instance, only 1 BS within a cluster of $(N^* + 1)$ BSs can be the master BS. Within the master BS, some UEs may obtain a better performance as a result of the scheduling process;
- All other $N^*$ BSs are slave BSs;
- Only the master BS's throughput performance is changed, the slave BSs' throughput remains the same;

An important remark is that for a network of $(N + 1)$ BSs, only a certain fraction of BSs can benefit from the proposed technique. The average number of BSs that can benefit is precisely $\frac{N+1}{N^*+1}$. In the performance comparison section, we will consider what the optimal cluster size is.

For each benefiting master BS, we considered UE pairs (2 UEs per pairing). For each pairing, a high capacity and a low capacity UE is selected. Typically, the high capacity UEs have a channel that is *at or near* capacity saturation. The expectation is that the saturated UE can receive additional interference allocated from the low capacity UE, such that the aggregate capacity is improved. The illustrations in Fig. 3 show the inter-BS clustering and intra-BS UE pairing schemes. In order to define a UE partnership (Fig. 3b), each high capacity UE with capacity $C(\gamma_\text{s})$ must define the degradation tolerance parameter $F$, where the capacity after scheduling is $\geqslant FC(\gamma_\text{s})$. Given this, a benefiting low capacity UE is optimally selected, such that the pair-wise aggregate capacity is maximised as a function of the slave BS cooperation cluster size $N^*$. The paper will now consider the potential performance gain for a *free-lunch* zero-degradation case ($F = 1$), and the more general case of allowable degradation ($F < 1$).

## IV. Performance Optimisation

### A. Problem Formulation

In the benefiting master BS, we consider a UE pairing that consists of two arbitrary UEs (Fig. 3b). One of the UEs has a high capacity channel (can be saturated) and the other has an unsaturated channel. The *problem formulation* has the following objective function and constraint:

- **Objective**: Maximise the sum-rate of the pairing;
- **Constraint**: Do not violate the degradation constraint ($F$) of the high capacity UE;

The case that the capacity of the unsaturated UE can be improved with no degradation to the high capacity UE, is the special *Free-Lunch (FL)* case. This can only occur if the high capacity UE's performance is already saturated. In the general case, the capacity of the unsaturated UE can be improved with some degradation to the saturated UE's capacity. Next, we examine both the special FL and the general case.

### B. SIR Approximation

First, we approximate the SIR expression given in Eq. (1), such that the SIR is tractably related to the number of interferers. By taking the expectation of the combined serving and interference BSs' multipath fading terms, a single term can be found (see Appendix B). Furthermore, it can be shown that the aggregate interference power is directly proportional to the number of interference terms (see Appendix C). By combining the results from Appendix B and C, the expectation of the SIR expression is:

$$\mathbb{E}[\gamma_{m,i,N}] = \frac{H d_{m,i}^{-\alpha}}{N\mathcal{D}(\Lambda, \alpha, N)}, \tag{3}$$

where:

$$H = e^{0.115\sigma^2 + 3}, \tag{4}$$

$$\mathcal{D}(\Lambda, \alpha, N) = \sum_{j=2}^{N+2} \frac{(\lambda\pi)^{\frac{\alpha}{2}} \Gamma(j - \frac{\alpha}{2})}{N\Gamma(j)}, \tag{5}$$

and $\Lambda$ is the BS density, $(N + 1)$ is the total number of BSs and $\Gamma(x) = \int_0^{+\infty} e^{-t} t^{x-1} \, \mathrm{d}t$.

*C. Free-Lunch (FL) Case*

In order to have a *free-lunch* solution, the sum-rate improvement in the unsaturated channel must come at a negligible capacity loss to the capacity of saturated channel. The paper defines:

- the saturated UE as $m = \text{s}$ and the unsaturated UE as $m = \text{u}$;
- $F = 1$ for free-lunch;
- the allocated number of interfering RRBs as $X$, which is upper-bounded by the difference between the control BS cluster size and 1 ($N^*$);

The paper also defines the reference system as a round-robin (RR) system with random uniform allocation of the RRBs. In such a system, the number of interference sources received by each UE is on average the same.

The free-lunch versus reference (FL,ref.) throughput gain is the UE pair sum-rate difference between prior to the IA procedure $C(\gamma_{\text{u},i,N})$ and after the allocation $C(\gamma_{\text{u},i,N-X})$:

$$\Delta C_{\text{FL,ref.}} = C(\gamma_{\text{u},i,N-X}) - C(\gamma_{\text{u},i,N}), \quad (6)$$
$$\text{subject to: } C(\gamma_{\text{s},i,N+X}) = FC(\gamma_{\text{s},i,N}).$$

The *maximum* amount of interference that can be allocated is (refer Appendix D for the proof):

$$X < \min\left(N\left\{\frac{\gamma_{\text{s},i,N}}{c\tan\left[\frac{FC(\gamma_{\text{s},i,N})}{a}\right] - b} - 1\right\}, N^*\right), \quad (7)$$

where in this free-lunch case, the constraint is $F = 1$. In practice, the amount of interference headroom which a saturated UE can tolerate is entirely determined by the reported SIR of the UE. As shown in Eq. (7), by knowing $\gamma_{\text{s},i,N}$, the value of $X$ can be estimated.

The advantage of the free-lunch solution is that no performance sacrifice is made towards any UE, whilst some UEs have enjoyed a significant sum-rate gain given by Eq. (6). The proposed scheme examines the average SIR achieved by UEs for a given UE location. Considering the majority of mobile data transfer is conducted by stationary or pedestrian UEs, we believe the average SIR will not significantly change over the course of a few seconds and therefore Transmission Time Interval (TTI) based synchronisation between BSs is not necessary.

*1) Master BS Results:* Fig. 4 presents the multi-BS and multi-UE simulation results, where the capacities achieved by the saturated and unsaturated UEs are plotted as a function of the amount of re-allocated interference RRBs ($X$), for different initial unsaturated channel conditions $\gamma_{\text{u},i,N}$. As the initial unsaturated channel strength decreases, the benefits of interference re-allocation diminish linearly. The key points are:

1) the capacity improvement to the unsaturated channels is significant for high values of re-allocation, with UE throughput gains of 80% achieved for a low $\gamma_{\text{u},i,N} = 1$ dB for $X$ varying from 0 to 18;
2) the capacity degradation to the saturated channel is negligible ($F > 0.99$).

Whilst not shown in Fig. 4, the *upper-bound* of the sum-rate gain is achieved under three conditions, namely: i) when the low capacity UE as an initial and minimum unsaturated capacity of $C(\gamma_{\text{u},i,N}) = 0$, ii) when the low capacity UE improves its capacity to the saturation point, and iii) under the condition that the saturated capacity UE remains saturated. In such a scenario, the resulting sum-rate gain is $C(\gamma_{s,i,N})$.

*2) Network Average Results:* The average sum-rate gain for the network take into account two averaging effects: i) the number of free-lunch UE pairs in the master BS, and ii) the number of master BSs in the network. Therefore, in order to translate UE pair improvements into network wide improvements, the former need to be discounted by the following:

$$\frac{P(\overline{\gamma} > \zeta)}{N^* + 1} = \frac{1}{(N^* + 1)(1 + \sqrt{\zeta}\arctan\sqrt{\zeta})}, \tag{8}$$

where $\zeta$ determines the saturation SIR. The value for $P(\overline{\gamma} > \zeta)$ can be found using averaged real network data or theory [26], [27]:

$$P(\overline{\gamma} > \zeta) = \frac{1}{1 + \sqrt{\zeta}\arctan\sqrt{\zeta}}, \tag{9}$$

for a pathloss exponent of $\alpha = 4$.

By considering the upper-bound sum-rate gain: for values of $C(\gamma_{s,i,N}) = 4$ bits/s/Hz, $N^* + 1 = 7$, $\zeta = 25$ dB, and $B = 20$ MHz bandwidth, the free-lunch sum-rate improvement averaged across the network is $\frac{P(\overline{\gamma}>\zeta)C(\gamma_{s,i,N})B}{N^*+1} = 0.41$ Mbits/s per BS. Although the averaged network improvements are modest, it is important to note that unlike other conventional schedulers (e.g., proportional fair), the free-lunch solution does not sacrifice any other metric of the UE performance (i.e., $F = 1$). We now show that throughput improvement gains can also be achieved if the saturated UE performance constraint in Eq. (6) is relaxed (i.e., $F < 1$).

*D. General Case: Non-Free-Lunch*

In the *non-free-lunch* (NFL) case, the problem formulation attempts to maximise the *sum-rate* of the UE pairing. The sum rate compared to the RR reference case is given by:

$$\Delta C_{\text{NFL,ref.}} = \\ C(\gamma_{u,i,N-X}) + C(\gamma_{s,i,N+X}) - C(\gamma_{u,i,N}) - C(\gamma_{s,i,N}). \tag{10}$$

Given the sum-rate, the paper attempts to find the optimal value of $X$ such that the sum-rate is maximised. It turns out that the optimal value of $X$ is:

$$X_{\text{opt}} = N^*, \tag{11}$$

as proven in Appendix D. That is to say, the greater the BS cluster size $(N^* + 1)$, the greater the sum-rate gain in both the master BS, and across the entire network. The resulting sum-rate gain is plotted in Fig. 5, as a factor of $F$ and the initial capacity of the unsaturated channel $C(\gamma_{u,i,N})$, for $X_{\text{opt}} = N^* = 18$. The results show that for a given UE pairing, the optimal free lunch factor $(F)$ that maximises the sum-rate gain can be found.

Now, consider the sum-rate gain as a function of the initial capacity of the unsaturated UE in Fig. 5. It is shown in Eq. (43) in Appendix D that the sum-rate gain is a monotonic function in terms of $C(\gamma_{u,i,N})$. This can be proven by taking the derivative of the sum-rate gain and showing it is strictly positive:

$$\frac{\mathrm{d}\Delta C_{\mathrm{NFL,ref.}}}{\mathrm{d}C(\gamma_{\mathrm{u},i,N})} = \frac{X}{N-X} \geqslant 0, \qquad (12)$$

conditioned on that the unsaturated UE is within the low SIR regime ($< 12$ dB). This can be observed in Fig. 5, whereby for the range of initial unsaturated UE capacity (0–1.5 bits/s/Hz), it has a monotonic relationship with the sum-rate gain. That is to say, provided that a UE pairing consists of a saturated and an unsaturated UE, the scheme will always benefit their sum-rate. This is true for any constraint value ($F$), however, it is worth noting that the overall sum-rate can be negative (degradation). Therefore, it is important to optimise $F$ correctly for a given UE-pairing. Optimally, the resulting sum-rate gain is up to 40 Mbits/s per cell over a 20 MHz bandwidth.

## V. COMPARATIVE PERFORMANCE STUDY OF DIFFERENT SCHEDULERS

The paper defines the following schedulers for comparison:

1) *Round Robin (RR)*: resource blocks are uniformly allocated across UEs and there is on average the same number of interference sources received by each UE;
2) *Proportional Fair (PF)*: resource blocks allocation is prioritised to UEs with poor signal quality [28], [29]. There is on average the same number of interference sources received by each UE;
3) *Interference Avoidance*: BSs allocate their resource blocks so that co-frequency interference between adjacent BSs are minimised [13];
4) *Interference Allocation (IA)*: interference is minimized to UEs with a *low* capacity through a UE pairing system. The optimal number of interfering resource blocks between UEs in a pairing is given by $X$, from Eq. (11). There is on average a different number of interference sources received by each UE, depending on their channel conditions;
5) *Alternative Scheme - Max SIR IA*: interference is minimized to UEs with a *high* capacity through a UE pairing system. The derivation is given in Appendix F, with a solution similar to the Water-filling power allocation solution. There is on average a different number of interference sources received by each UE, depending on their channel conditions.

The simulation results consider a multi-BS environment, whereby the performance of UEs in a single master BS's are sampled. Fig. 6 shows the mean achieved throughput for different *unfairness ratio* values. The unfairness ratio is defined as the ratio between the maximum and minimum SIR of UEs in a BS:

$$\Theta = 10\log_{10}\left(\frac{\gamma_{\max}}{\gamma_{\min}}\right) \quad \mathrm{dB}. \qquad (13)$$

Therefore, the greater the ratio, the greater the discrepancy between the UEs' performances.

The results show that the proposed *IA* scheduler:

- achieves a uniform BS throughput performance for any multi-UE channel quality by taking into account the capacity saturation levels.
- achieves significantly better BS throughput than RR, PF, interference avoidance, max SIR allocation schedulers (up to 80%), especially in the case of a high discrepancy between the DL SIR levels in a BS (high $\Theta$). This can occur in urban environments, where the coverage in alley-ways and indoor areas can be very poor.

From the results in Fig. 6, it can be seen that for low to medium values of $\Theta$ ($< 20$ dB) range, the IA scheme is just as good as the PF scheme. The time-division based interference avoidance scheme in [13] also performances quite well. At medium-to-high $\Theta$ ($> 25$ dB), the proposed scheme maintains the same aggregate throughput performance, whilst the other schemes sacrifice performance of one UE for the improvement of another. The proposed scheduler avoids sacrifice by allocating interference in such a way that it maintains a constant high BS throughput for a wide range of channel conditions. The caveat is that in IA, only the master BS benefits, out of a cluster of $(N^* + 1)$ BSs.

In summary, in the cells that benefit from the proposed scheme, the results show that the throughput improvement is up to 28 Mbits/s (67%) better than alternative schedulers, especially when there is a high SNR discrepancy in the cell. These large discrepancies are typically found in urban cell coverage zones, where there is up to 25–30 dBs of difference in UE SNRs, as shown in [30].

## VI. Green Multimedia Delivery

In this section, the paper examines the energy saving and multimedia QoE results that arise from the proposed scheduler. As shown in the previous sections, the proposed scheduler can offer dramatic pair-wise rate improvements to UEs with a high SIR discrepancy. Given that not all UEs in a cell use multimedia services, thus UE pairing should be targeted to those who do. In particular, the unsaturated UEs who use multimedia services should be paired with saturated ones and their interference reallocated so that their service quality (QoE) is improved. Thus with some additional multimedia service management combined with the proposed resource allocation policy, we might significantly improve the QoE of unsaturated UEs by shifting interference away from them.

### A. Green Multimedia Performance Metrics

*1) Energy Consumption Metric:* The instantaneous total power consumption of a BS can be split between the transmission (radio-head, RH) and the static (over-head, OH) part [4], [31]:

$$P_{\text{total}} = P_{\text{RH}} + P_{\text{OH}}, \tag{14}$$

where $P_{\text{RH}}$ is the ratio between the transmit power $P$ and the radio-head amplifier efficiency $\mu$. In fact, the full energy consumption will also include embodied and backhaul consumption values, which make up approximately 35% of the total expenditure [4].

In order to derive the energy consumption, a notion of transmission time is introduced. The paper defines the average traffic load of the BS as the ratio between traffic demand $L$ and BS throughput $C$. Therefore, given that the BS transmits for duration $T_{\text{TX}}$ when there is traffic and consumes static energy constantly, the total energy consumption over time $T_{\text{total}}$ can be written as [32]:

$$\begin{aligned} E_{\text{total}} &= P_{\text{RH}} T_{\text{TX}} + P_{\text{OH}} T_{\text{total}}, \\ &= T_{\text{total}} \left( \frac{P}{\mu} \frac{C_{\text{traffic}}}{C} + P_{\text{OH}} \right). \end{aligned} \tag{15}$$

Normalising against the total time means that the normalised consumption with respect to a certain traffic rate and BS throughput is: $P_{\text{RH}}\frac{C_{\text{traffic}}}{C} + P_{\text{OH}}$. Values for the power consumption can be found in [31].

*2) Quality of Experience Metric:* Evaluation and performance of Quality of Experience (QoE) in mobile cellular systems is currently very active research area. Across the literature, various relationships have been derived that link QoE with the major transmission system and video codec parameters. In this paper, we apply recently proposed QoE models for H.264 Scalable Video Coder (H.264/SVC) services [33] that provide analytical relations between the average data rates, $R(q,t)$, and average subjective video quality based on Mean Opinion Score (MOS), $Q(q,t)$, as a function of major H.264/SVC parameters: the frame rate ($t$, Hz) and the quantisation step-size ($q$). The quantisation step size is a dimensionless integer that used for quantising DCT coefficients after transform. By increasing the value, DCT coefficients are increasingly approximated with rougher granularity.

In particular, from [33] we have that:

$$R(q,t) = R_{\max} \left(\frac{q}{q_{\min}}\right)^{-a'} \left(\frac{t}{t_{\max}}\right)^{-b'}, \quad (16)$$

$$Q(q,t) = Q_{\max} \frac{\exp\left[-c'\left(\frac{q}{q_{\min}}\right)\right]}{\exp(-c')} \frac{1-\exp\left[-d'\left(\frac{t}{t_{\max}}\right)\right]}{1-\exp(-d')}, \quad (17)$$

where $q_{\min} = 16$ and $t_{\max} = 30$ represent the minimum quantisation step-size, and maximum frame rate (frames per second) parameters used by the video codec that result in the maximum possible data rate $R(q,t) = R_{\max} = 806$ kbps. Without loss of generality, in this paper we apply content-dependent parameters $a' = 1.149$, $b' = 0.577$, $c' = 0.12$ and $d' = 8.24$ derived for *Foreman* video sequence. The numerical model parameters used describe how fast the achieved bit rate is reduced by increasing the quantisation step-size and/or reducing the frame rate. From the parametric model described above, we are able to (implicitly) derive QoE-parameter $Q$ dependence on available throughput $R$ which we use in the system evaluation.

## B. Green Multimedia Delivery Simulation Results

*1) Energy Saving Performance:* The simulation results in the rest of this section consider an arbitrary UE pairing, whereby one UE is saturated in capacity with SIR ($\gamma_{s,i,N}$), and one UE is unsaturated with SIR ($\gamma_{u,i,N}$). From the results in Fig. 4, the resulting UE-pair energy consumption benefit from employing the proposed scheduler can be found using Eq. (15). For power consumption values of $P = 40$ W, $\mu = 0.3$, and $P_{\text{OH}} = 120$ W, the free-lunch energy consumption saving is upper-bounded to (see Appendix E for the proof):

$$E_\downarrow = \frac{1}{2\left[1 + \left(\frac{C(\gamma_s)}{C_{\text{traffic}}}\right)\frac{P_{\text{OH}}}{P_{\text{RH}}}\right]}. \quad (18)$$

The energy saving upper-bound is given as $1/2[1 + \frac{P_{\text{OH}}}{P_{\text{RH}}}]$, which is approximately 25%, assuming the power consumption values given in [31].

The results in Fig. 7 present the case for a common saturated UE ($\gamma_{s,i,N} = 25$ dB) paired with various unsaturated UEs each with SIR ($\gamma_{u,i,N}$) varying from 1 dB to 10 dB. The results show that as the level of interference allocation (spreading) increases, the energy saving for the UE pair improves significantly. For a medium discrepancy between the saturated and unsaturated UEs ($\Theta > 20$ dB) the IA scheme is largely ineffective until it reaches high values. This is due to the nature of the capacity profile for realistic modulation-and-coding schemes, whereby the capacity improvement is small at low SIRs ($\gamma_{u,i,N} < 5$ dB). For unsaturated UEs with higher unsaturated SIRs ($\gamma_{u,i,N} \geqslant 10$ dB), the energy saving improves more readily. The energy saving or capacity improvement is closely related to the rate of increase (derivative) of the combined MCS curve. At high SNRs, the derivative is smaller and this is why the energy saving for the $\gamma_{u,i,18} = 10$ dB case is smaller than its counter-parts in Fig. 7.

In the case of full interference RRB reallocation ($X = 18$), the energy saving is approximately 20%. The fundamental limit of the energy saving in any radio resource management scheme (RRM) is given by (18), which is approximately 25%.

*2) QoE Performance:* Previously, the paper showed how in the cells that benefit from the scheme, the throughput improvement is up to 28 Mbits/s (67%), and the energy reduction is 20% better than alternative schedulers, especially when there is a high SNR discrepancy. As discussed, this typically occurs in urban cellular coverage areas [30]. In this sub-section, we consider the QoE performance in terms of the viewing perception of mobile video streaming, which is previously defined as the mean opinion score (MOS). The MOS metric is derived from achievable rates using parameters for H.264/SVC compressed *Foreman* video sequence and Eq. (16).

The results in Fig. 8 consider a pair of UEs. The saturated capacity UE has a full MOS performance of 4.6 ($\gamma_{s,i,N} = 25$ dB). The other UE in the pairing has an unsaturated capacity performance ($\gamma_{u,i,N} = \{1, 4, 7, 10\}$ dB). Therefore, the discrepancy or the unfairness ratio in this pairing is $\Theta = \{24, 21, 18, 15\}$ dB. It can be seen that the greater the discrepancy ($\Theta$) or the lower the unsaturated UE's SIR ($\gamma_{u,i,N}$), the greater the potential for MOS improvement. The results in Fig. 8 show that as the level of interference spreading increases, the MOS for the UE pair improves. For a UE with a low SIR ($\gamma_{u,i,N} = 1$ dB) and medium-high discrepancy ($\Theta = 24$ dB), the improvement can be up to 20%. For an unsaturated UE with a medium SIR ($\gamma_{u,i,N} = 10$ dB) and medium-low discrepancy ($\Theta > 15$ dB), the improvement can be up to 7%. Clearly for unsaturated UEs with a higher SIR than 10 dB ($\Theta < 15$ dB), the MOS improvement diminishes to below 7%. Therefore, the scheme primarily benefits UEs with a poor SIR performance.

The caveat with this scheme is that it requires cooperation from a cluster of BSs. The time averaged network-wide improvement of multiple BSs assisting a few UEs is approximately 1%, for $N^* = 18$ BSs. Further research should focus on how to optimise BS cluster sizes to take into account the role of dominant interference sources for specific UEs.

## VII. CONCLUSIONS

This paper has proposed a novel scheduler that allocates interference from unsaturated to saturated capacity channels. The investigation has shown that the proposed scheduler offers attractive performance improvements in cell throughput (67%), energy expenditure (20%), and mean-opinion-score (MOS) in video streaming (7–20%). This is particularly the case in cells with a high discrepancy in SNR values, typically found in urban areas. The detailed analysis shows that a *free-lunch* solution for the proposed scheduler can be achieved, whereby a UE pair sum-rate gain can be achieved for some channels at no loss to other channels. Furthermore, by relaxing the constraints, the *non-free-lunch* solution can yield a greater multi-UE throughput gain. The paper presents the simulated results, as well as a closed-form theoretical formula for the proposed scheduler and compares it with conventional scheduling techniques that do not have knowledge of capacity saturation. The results show that for the targeted UE pairing, the IA scheduler can improve throughput by up to 80% with negligible side-effects to other UEs. The caveat of the proposed scheme is that only a small fraction of BSs and UEs can benefit from this scheme at any time, and future work will examine optimising the cell cluster size.

## APPENDIX A
## ARCTANGENT CAPACITY EXPRESSION

The arctangent capacity function that takes into account the capacity saturation level, modulation-and-coding rate, is given by [18]:

$$C(\gamma) \approx a \arctan(\gamma, b, c) = a \arctan\left(\frac{\gamma + b}{c}\right), \quad (19)$$

where the SNR ($\gamma$) is in linear term, given by $\gamma = \frac{|h|^2 P}{N_0}$. The parameters $a$, $b$ and $c$ are adjustment factors, whereby through curve-fitting, an empirical relationship can be established with the modulation bits/symbol rate ($\mathcal{N}$) and FEC coding rate ($\mathcal{R}$) of MCS:

$$\begin{cases} a = -0.03 + 0.015\mathcal{N} + 0.4\mathcal{N}\mathcal{R} - 0.001\mathcal{N}^2 + 0.04\mathcal{R}^2 \\ b = 2.1 - 1.1\mathcal{N} - 2.6\mathcal{R} + 1.46\mathcal{N}\mathcal{R} + 0.046\mathcal{N}^2 - 0.5\mathcal{R}^2 \\ c = 17.8 - 10\mathcal{N} - 25.3\mathcal{R} + 14.54\mathcal{N}\mathcal{R} + 0.54\mathcal{N}^2 - 3.3\mathcal{R}^2 \end{cases}. \quad (20)$$

The *combined adaptive MCS* for LTE in a 3GPP channel has a unique set of adjustment parameters: $a = 2.27$, $b = 13$, and $c = 40$. Based on the simulated adaptive MCS of the LTE physical layer in [18], the theoretical Eq. (2) with different modulation and coding rates is presented in Fig. 9, along with the combined theoretical MCS curve. The approximation accuracy is very high, with any given MCS achieving: a mean error of $2.5 \times 10^{-2}$ bit/s/Hz across all MCSs, and a variance of $2.8 \times 10^{-3}$.

## APPENDIX B
## AGGREGATE FADING EXPRESSION

The multipath and log-normal shadow fading can be combined into a modified log-normal distributed with probability density function (PDF) [34], [35]:

$$f_H(s; \widetilde{\sigma}, \widetilde{\mu}) = \frac{1}{s\widetilde{\sigma}\sqrt{2\pi}} e^{-\frac{(\ln s - \widetilde{\mu})^2}{2\widetilde{\sigma}^2}}, \quad (21)$$

where the modified values are $\widetilde{\mu} = -0.58$ and $\widetilde{\sigma}^2 = 0.23(\sigma^2 + 5.57^2)$ [35]. The mean of the combined multi-path and shadow fading distribution is given as: $H = e^{0.115\sigma^2 + 3}$.

# APPENDIX C

## AGGREGATE INTERFERENCE EXPRESSION

The paper defines $\gamma$ as the Signal-to-Interference Ratio (SIR), where the SIR of a UE $m$, attached to BS $i$, with $N$ interference RRBs is defined as:

$$\gamma_{m,i,N} = \frac{P_{m,i}\lambda d_{m,i}^{-\alpha} e^{0.115\sigma^2 + 3}}{\sum_{j=1, j\neq i}^{N+1} P_{m,j}\lambda d_{m,j}^{-\alpha}}, \tag{22}$$

after substituting in the combined fading distributions Eq. (21). The combined fading and shadowing terms in the interference are assumed to not significantly affect the aggregated value [36].

As stated in [26], the distribution of the distance from the origin to the nearest serving BS $d_{m,i}$ follows a Rayleigh distribution: $D_{m,i} \sim \text{Rayleigh}\left(\frac{1}{\sqrt{2\Lambda\pi}}\right)$, where $\Lambda$ denotes the BS density and its relationship to inter-site distance of $\delta$ is $\Lambda \propto 1/\delta^2$. However, the extension to the second, third and $n^{\text{th}}$ nearest BS has not been considered in [26]. The paper now provides the derivation of the distribution of $d_{m,j}$ (the distance from $j^{\text{th}}$ interfering BS to the origin, which is equivalent to the distance from $(j+1)^{\text{th}}$ nearest BS to the origin). The approach of the derivation continues the same fashion of the derivation of $d_{m,i}$. Recall that the probability of no BS is closer than r is given by

$$\mathbb{P}(0, \mathsf{r}) = e^{-\Lambda\pi\mathsf{r}^2}. \tag{23}$$

Hence, the probability of finding at least one BS within the distance of r is

$$\mathbb{P}(\mathsf{n} \geqslant 1, \mathsf{r}) = 1 - e^{-\Lambda\pi\mathsf{r}^2}. \tag{24}$$

The probability of finding exactly one BS within the distance of r is

$$\mathbb{P}(1, \mathsf{r}) = \Lambda\pi\mathsf{r}^2 e^{-\Lambda\pi\mathsf{r}^2}. \tag{25}$$

Therefore, the probability of finding at least two BSs within the distance of r is

$$\begin{aligned}\mathbb{P}(\mathsf{n} \geqslant 2, \mathsf{r}) &= 1 - [\mathbb{P}(0, \mathsf{r}) + \mathbb{P}(1, \mathsf{r})], \\ &= 1 - \left[e^{-\Lambda\pi\mathsf{r}^2} + \Lambda\pi\mathsf{r}^2 e^{-\Lambda\pi\mathsf{r}^2}\right]. \end{aligned} \tag{26}$$

Continuing in a similar manner and the probability of finding at least j BSs within the distance of r is given by

$$\begin{aligned}\mathbb{P}(\mathsf{n} \geqslant \mathsf{j}, \mathsf{r}) &= 1 - \sum_{\mathsf{m}=0}^{\mathsf{j}-1} \mathbb{P}(\mathsf{m}, \mathsf{r}), \\ &= 1 - \left[e^{-\Lambda\pi\mathsf{r}^2} + \Lambda\pi\mathsf{r}^2 e^{-\Lambda\pi\mathsf{r}^2} \cdots + \frac{(\Lambda\pi\mathsf{r}^2)^{\mathsf{j}-1}}{(\mathsf{j}-1)!} e^{-\Lambda\pi\mathsf{r}^2}\right]. \end{aligned} \tag{27}$$

The probability that the $j^{\text{th}}$ nearest BS to the origin is found in the annulus between the concentric circles radii r and $r + \Delta r$, which is the difference between $\mathbb{P}(\mathsf{n} \geqslant \mathsf{j}, \mathsf{r} + \Delta\mathsf{r})$ and $\mathbb{P}(\mathsf{n} \geqslant \mathsf{j}, \mathsf{r})$ and is expressed as

$$\mathbb{P}(\mathsf{j}, \mathsf{r} \sim \mathsf{r} + \Delta\mathsf{r}) = \mathbb{P}(\mathsf{n} \geqslant \mathsf{j}, \mathsf{r} + \Delta\mathsf{r}) - \mathbb{P}(\mathsf{n} \geqslant \mathsf{j}, \mathsf{r}). \tag{28}$$

The PDF of the distance between the j$^{\text{th}}$ nearest BS to the origin is obtained by letting $\Delta r$ approach to the infinitesimal interval $dr$ and then differentiating Eq. (28) with the reference of Eq. (27)

$$f_{R_j}(r;j) = \frac{d\mathbb{P}(n \geqslant j, r)}{dr} = \frac{2(\Lambda\pi)^j}{(j-1)!} r^{2j-1} e^{-\Lambda\pi r^2}, \quad j \geqslant 1. \tag{29}$$

For $j = 1$, Eq. (29) reduces to $2\Lambda\pi r e^{-\Lambda\pi r^2}$ which is the same Rayleigh distribution of the distance to the nearest serving BS derived in [26]. Setting $y = \Lambda\pi r^2$, Eq. (29) can be re-rewritten as

$$\begin{aligned} f_{R_j}(y;j) &= \frac{d\mathbb{P}(n \geqslant j, r)}{dy} = \frac{d\mathbb{P}(n \geqslant j, r)}{2\Lambda\pi r dr}, \\ &= \frac{(\Lambda\pi)^{j-1}}{(j-1)!} r^{2j-2} e^{-\Lambda\pi r^2} = \frac{y^{j-1}}{(j-1)!} e^{-y}, \quad j \geqslant 1, \end{aligned} \tag{30}$$

which gives the classic form of gamma distribution with shape parameter $j$ and scale parameter $1$ $\left(\Lambda\pi R_j^2 \sim \Gamma(j, 1)\right)$. Furthermore, setting $z = 2\Lambda\pi r^2$, $z$ can be proved to follow a chi-squared distribution with $2j$ degrees of freedom $\left(2\Lambda\pi R_j^2 \sim \chi^2(2j)\right)$

$$\begin{aligned} f_{R_j}(z;j) &= \frac{d\mathbb{P}(n \geqslant j, r)}{dz} = \frac{d\mathbb{P}(n \geqslant j, r)}{4\Lambda\pi r dr}, \\ &= \frac{(2\Lambda\pi)^{j-1}}{2^j(j-1)!} r^{2j-2} e^{-\Lambda\pi r^2} = \frac{z^{j-1}}{2^j(j-1)!} e^{-\frac{z}{2}}, \quad j \geqslant 1, \end{aligned} \tag{31}$$

For a given arbitrary UE at a distance $d_{m,i}$ from the serving BS, the next challenge is how to find the distribution of the interference signal power, which is related to $d_{m,j}^{-\alpha}$. Let the random variable $\mathfrak{D}_{m,j} = D_{m,j}^{-\alpha}$, the expected value of $\mathfrak{D}_{m,j}$ by definition is given by:

$$\begin{aligned} \mathbb{E}[\mathfrak{D}_{m,j}] &= \frac{\sum_{j=2}^{N+2} \int_0^{+\infty} r^{-\alpha} \frac{2(\Lambda\pi)^j}{(j-1)!} r^{2j-1} e^{-\Lambda\pi r^2} \, dr}{N}, \\ &= \sum_{j=2}^{N+2} \frac{(\lambda\pi)^{\frac{\alpha}{2}} \Gamma(j - \frac{\alpha}{2})}{N\Gamma(j)} = \mathcal{D}(\Lambda, \alpha, N), \end{aligned} \tag{32}$$

where $\Gamma(x) = \int_0^{+\infty} e^{-t} t^{x-1} \, dt$.

Assuming all BSs have the same transmit power $P$, the resulting expected SIR is therefore:

$$\mathbb{E}[\gamma_{m,i,N}] \approx \frac{HP\lambda d_{m,i}^{-\alpha}}{NI} = \frac{Hd_{m,i}^{-\alpha}}{N\mathcal{D}(\Lambda, \alpha, N)}, \tag{33}$$

where the average interference power $I = P\lambda\mathcal{D}$. It is worth stating that the interference power from each BS is not the same (as the distances vary). However, the aggregate interference power from all BSs is expressed as an expectation ($I$) in Eq. (33). Therefore, there is a linear scaling relationship between the aggregate interference power ($NI$) and the number of BSs, under the condition of BS density and some other factors like the UE-BS distance distribution.

## APPENDIX D

### INTERFERENCE ALLOCATION

In the *free-lunch* case, the problem formulation attempts to find the allocation factor ($X$) that achieves a free-lunch factor of ($F$):

$$C(\gamma_{\text{s},i,N+X}) > FC(\gamma_{\text{s},i,N}),$$
$$\gamma_{\text{s},i,N+X} > c\left\{\tan\left[\frac{FC(\gamma_{\text{s},i,N})}{a}\right]\right\} - b, \quad (34)$$
$$X(\gamma, F, N) < N\left\{\frac{\gamma_{\text{s},i,N}}{c\tan\left[\frac{FC(\gamma_{\text{s},i,N})}{a}\right] - b} - 1\right\}.$$

The number of interference resources allocated cannot exceed the total co-frequency resources available ($N^*$). Therefore, the process of finding $X(\gamma, F, N)$ comes from upper-bounding $X(\gamma, F, N)$ in Eq. (34) by the value $N^*$:

$$X(\gamma, F, N)$$
$$< \min\left(N\left\{\frac{\gamma_{\text{s},i,N}}{c\tan\left[\frac{FC(\gamma_{\text{s},i,N})}{a}\right] - b} - 1\right\}, N^*\right). \quad (35)$$

In the general case, the problem formulation attempts to maximise the sum-rate in the multi-UE BS. Assuming that the unsaturated channel has a small SIR, the sum-rate gain can be reformulated to:

$$\Delta C_{\text{NFL,ref.}} = C(\gamma_{\text{u},i,N-X}) + C(\gamma_{\text{s},i,N+X}) \\ - C(\gamma_{\text{u},i,N}) - C(\gamma_{\text{s},i,N}). \quad (36)$$

The last two terms of Eq. (36) are not functions of $X$. Hence, it is sufficient to maximise the first two terms in order to maximise the sum-rate. Without loss of generality, $C(\gamma_{\text{u},i,N-X})$ and $C(\gamma_{\text{s},i,N+X})$ can be approximated as:

$$C(\gamma_{\text{u},i,N-X}) = \frac{a}{c}\left(\frac{N}{N-X}\gamma_{\text{u},i,N} + b\right), \quad (37)$$
$$C(\gamma_{\text{s},i,N+X}) = \frac{a\pi}{2} - \frac{ac(N+X)}{N\gamma_{\text{s},i,N} + b(N+X)}, \quad (38)$$

where the approximations of the arctangent function ($\arctan(x) \approx x$ when $x$ is a small number and $\arctan(x) \approx \frac{\pi}{2} - \frac{1}{x}$ when $x$ takes large values) have been applied. Taking the derivative of the sum of Eq. (36) with respect to $X$ gives:

$$\frac{d\Delta C_{\text{NFL,ref.}}}{dX} = aN\left\{\frac{\gamma_{\text{u},i,N}}{c(N-X)^2} - \frac{c}{[N + \frac{b}{\gamma_{\text{s},i,N}}(N+X)]^2}\right\}. \quad (39)$$

By setting Eq. (39) equal to zero, two critical points $X_{\text{c1}}$ and $X_{\text{c2}}$ of Eq. (36) can be found as follows:

$$X_{\text{c1}} = \frac{N(c\sqrt{\gamma_{\text{s},i,N}} + b\sqrt{\gamma_{\text{u},i,N}} + \sqrt{\gamma_{\text{s},i,N}}\sqrt{\gamma_{\text{u},i,N}})}{c\sqrt{\gamma_{\text{s},i,N}} - b\sqrt{\gamma_{\text{u},i,N}}}, \quad (40)$$

$$X_{\text{c2}} = \frac{N(c\sqrt{\gamma_{\text{s},i,N}} - b\sqrt{\gamma_{\text{u},i,N}} - \sqrt{\gamma_{\text{s},i,N}}\sqrt{\gamma_{\text{u},i,N}})}{c\sqrt{\gamma_{\text{s},i,N}} + b\sqrt{\gamma_{\text{u},i,N}}}. \quad (41)$$

Moreover, by observing Eq. (39), the third critical point $X_{c3}$ is equal to $N$, when the derivative of Eq. (36) does not exist. Substituting $X_{c1}$, $X_{c2}$ and $X_{c3}$ in Eq. (36), the sum-rate gain can be found to be maximised when $X = X_{c3} = N$. Given that the cluster size of controllable interference sources is $N^* + 1 \leqslant N + 1$, the optimal $X$ is the difference between the cluster size and 1 ($N^*$). This conclusion holds for the maximum sum-rate gain of a single master BS, and the sum-rate gain averaged across the whole network.

Now, consider the sum-rate gain as a function of the initial capacity of the unsaturated UE. From Eq. (37), $C(\gamma_{\mathrm{u},i,N-X})$ can be expressed in terms of $C(\gamma_{\mathrm{u},i,N})$:

$$C(\gamma_{\mathrm{u},i,N-X}) = \frac{N}{N-X} C(\gamma_{\mathrm{u},i,N}) - \frac{abN}{c(N-X)}. \tag{42}$$

Taking the derivative of Eq. (36) with respect to $C(\gamma_{\mathrm{u},i,N})$ gives:

$$\frac{\mathrm{d}\Delta C_{\mathrm{NFL,ref.}}}{\mathrm{d}C(\gamma_{\mathrm{u},i,N})} = \frac{X}{N-X} \geqslant 0. \tag{43}$$

The sum-rate gain is therefore a monotonic function in terms of $C(\gamma_{\mathrm{u},i,N})$ conditioned on the fact that the unsaturated UE is within the low SIR regime so that Eq. (37) holds.

## APPENDIX E

## ENERGY SAVING

For a pair of saturated and unsaturated UEs, the highest achievable capacity gain is for the unsaturated UE (with negligible capacity, $C(\gamma_{\mathrm{u}}) \approx 0$) to rise to saturation and for the other saturated UE to remain saturated. In that case, the energy consumed by the serving-BS is reduced by:

$$\begin{aligned} E_{\downarrow} &= \frac{P_{\mathrm{RH}}\frac{C_{\mathrm{traffic}}}{C(\gamma_{\mathrm{s}})+C(\gamma_{\mathrm{u}})} + P_{\mathrm{OH}} - P_{\mathrm{RH}}\frac{C_{\mathrm{traffic}}}{2C(\gamma_{\mathrm{s}}))} - P_{\mathrm{OH}}}{P_{\mathrm{RH}}\frac{C_{\mathrm{traffic}}}{C(\gamma_{\mathrm{s}})+C(\gamma_{\mathrm{u}})} + P_{\mathrm{OH}}}, \\ &= \frac{1}{2\left[1 + \left(\frac{C(\gamma_{\mathrm{s}})}{C_{\mathrm{traffic}}}\right)\frac{P_{\mathrm{OH}}}{P_{\mathrm{RH}}}\right]}. \end{aligned} \tag{44}$$

## APPENDIX F

## MAX SIR INTERFERENCE ALLOCATION

By employing the unsaturated capacity expression [21], the optimal level of interference allocation $N_m^{\dagger}$ for each UE $m$ is derived using Lagrangian optimisation with a Lagrangian constant $\nu$. The interference allocation constraint is:

$$\sum_{m=1}^{M} N_m = N^*, \tag{45}$$

where the constraint is that the total number of interfering RRBs for $M = 2$ UEs is the difference of the number of interfering cells and 1 in the cell cluster (size $N^*$).

The resulting Lagrangian is:

$$\mathcal{L} = \sum_{m=1}^{M=2} \log_2\left(1 + \frac{Hd_m^{-\alpha}}{N_m\mathcal{D}}\right) - \nu\sum_{m=1}^{M} N_m, \tag{46}$$

where $\mathcal{D}$ is closely related to the expected interference power. The derivative of the Lagrangian is:

$$\frac{\partial \mathcal{L}}{\partial N_m} = -\frac{\frac{H d_m^{-\alpha}}{\mathcal{D}}}{\ln(2) N_m \left(N_m + \frac{H d_m^{-\alpha}}{\mathcal{D}}\right)} - \nu. \tag{47}$$

By maximising the Lagrangian with respect to $N_m$, the result is:

$$N_m^* = \frac{1}{2}\sqrt{A_m \left(A_m - \frac{4}{\nu \log(2)}\right)} - \frac{A_m}{2}, \tag{48}$$

where $A_m = \frac{H d_m^{-\alpha}}{\mathcal{D}}$. One needs to choose a suitable Lagrangian multiplier $\nu$ such that Eq. (48) meets the constraint in Eq. (45). This is effectively a maximum SIR solution, whereby the channel with the strongest SIR received the least level of interfering RRBs.

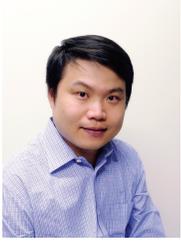

**Siyi Wang** received his Ph.D. degree in Electronic and Electrical Engineering from the University of Sheffield, UK, in 2014. He is currently a researcher in the Institute for Telecommunications Research at the University of South Australia. His research interests include: indoor-outdoor network interaction, small cell deployment, stochastic geometry, theoretical frameworks for complex networks and nano-particle communications.

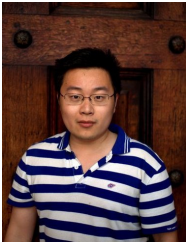

**Weisi Guo** received his M.Eng., M.A. and Ph.D. degrees from the University of Cambridge. He is currently an Assistant Professor at the University of Warwick, and is the author of the VCESIM LTE System Simulator. He is affiliated with New York University's Centre for Urban Science and Progress (CUSP). He has published over 50 peer-reviewed papers in the areas of energy/cost-efficiency of 4G networks, multi-user cooperative transmission, nano-particle communications, and mobile sensing.

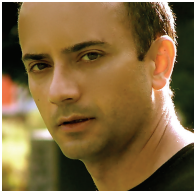

**Chadi Khirallah** received Ph.D. degree from the University of Lancaster, U.K, 2006. He was an Assistant Researcher from 2007–10 at the University of Strathclyde, then a Senior Research Fellow at the University of Edinburgh from 2010-13. Dr. Khirallah has joined NEC from 2013 as a Senior LTE-A Architecture and Protocols Engineer. His interests include LTE-A and 5G, multimedia applications, energy efficient architectures and deployment techniques in green cellular networks, and heterogeneous networks.

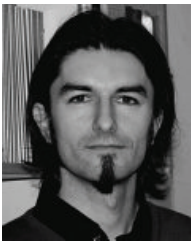

**Dejan Vukobratović** received Dr.-Ing. degree from the University of Novi Sad, Serbia, 2008. Since then he has been an Assistant Professor at the University of Novi Sad. In 2010, he was a Marie Curie Intra European Fellow at the University of Strathclyde. From 2011–14 his research is supported by Marie Curie European Reintegration Grant. His interests include sparse-graph codes, iterative decoding and network coding with applications in multimedia communications and wireless sensor networks.

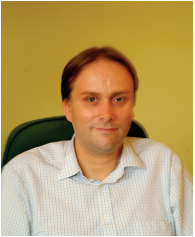 **John Thompson** currently holds a personal chair in Signal Processing and Communications at the School of Engineering in the University of Edinburgh. He specializes in antenna array processing, cooperative communications systems and energy efficient wireless communications. His work in these areas is highly cited and his h-index is currently 18. He is an elected Member-at-Large for the Board of Governors of the IEEE Communications Society from 2012-14. He was a technical programme co-chair for the 2013 IEEE Vehicular Technology Spring Conference and serves as a track chair on Green Communications for the 2014 IEEE International Conference on Communications.

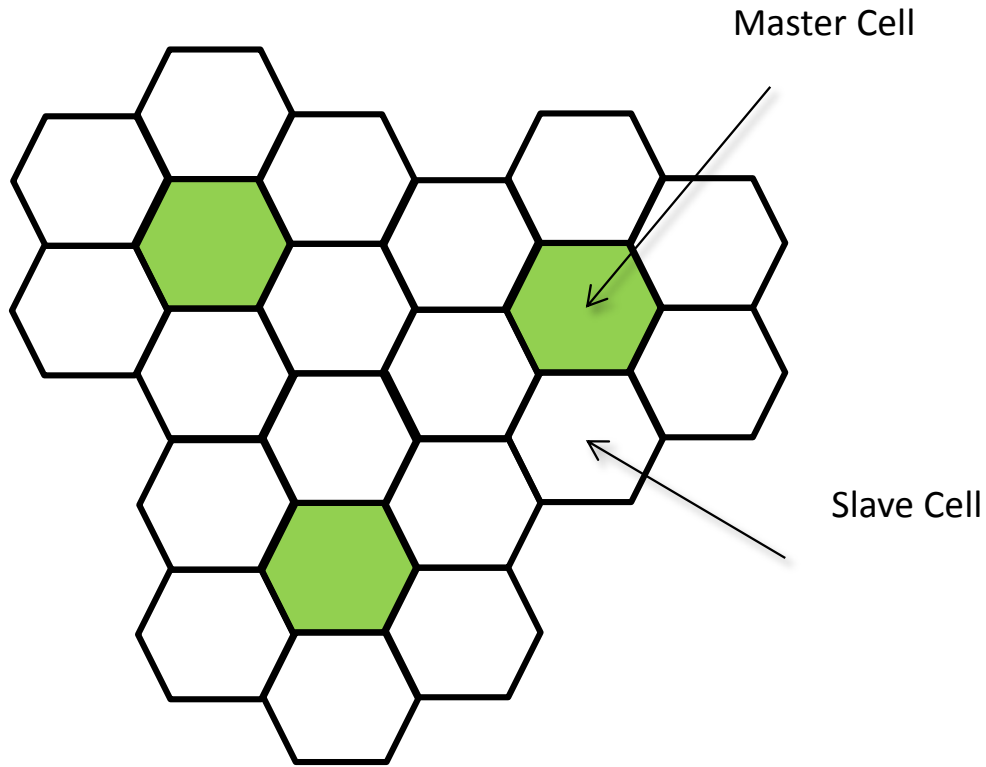

**a) Cell Clustering and Inter-Cell Coordination Scheme**

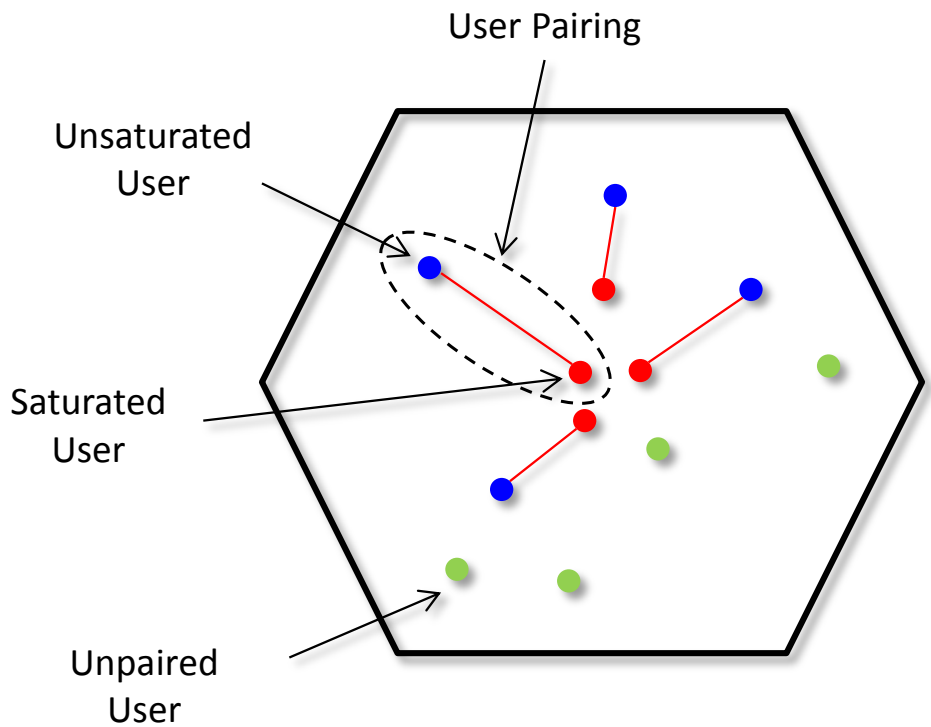

**b) Intra-Cell Pairing Scheme**

Fig. 3. Illustration of: a) inter-BS coordination, and b) intra-BS pairing schemes.

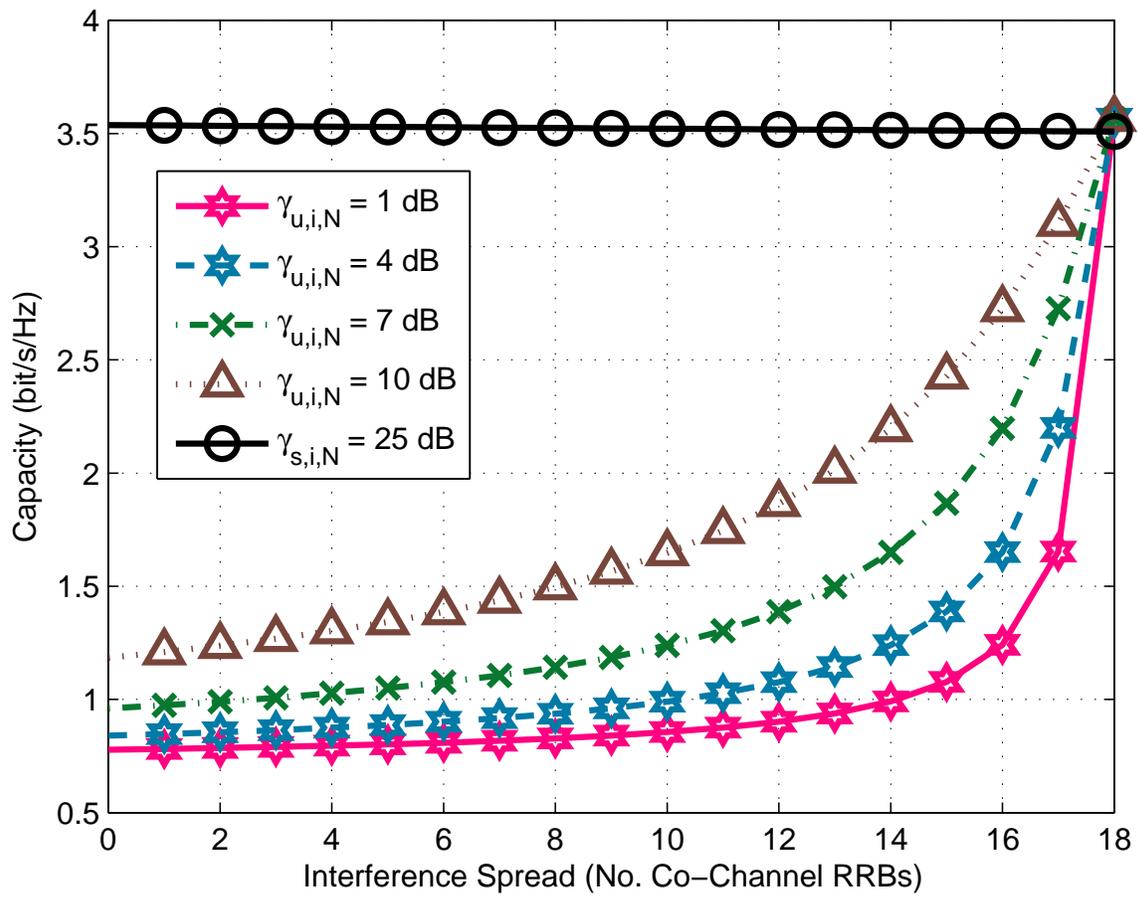

Fig. 4. Simulated capacity of saturated and unsaturated UE channels with different levels of interference allocation $X$.

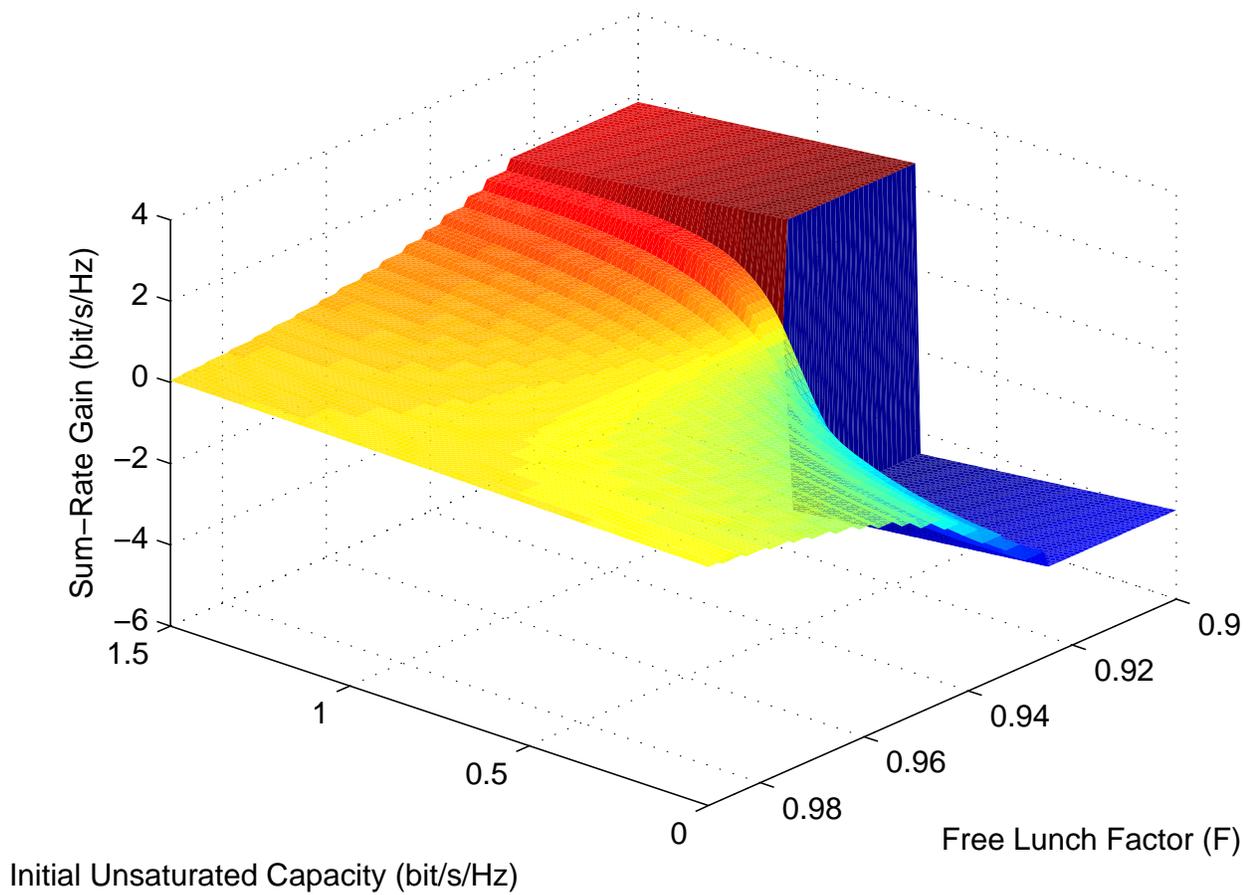

Fig. 5. Simulated sum-rate gain of IA compared to RR reference, as a function of the free-lunch factor ($F$) and the initial capacity of the unsaturated channel $C(\gamma_{\mathrm{u},i,N})$.

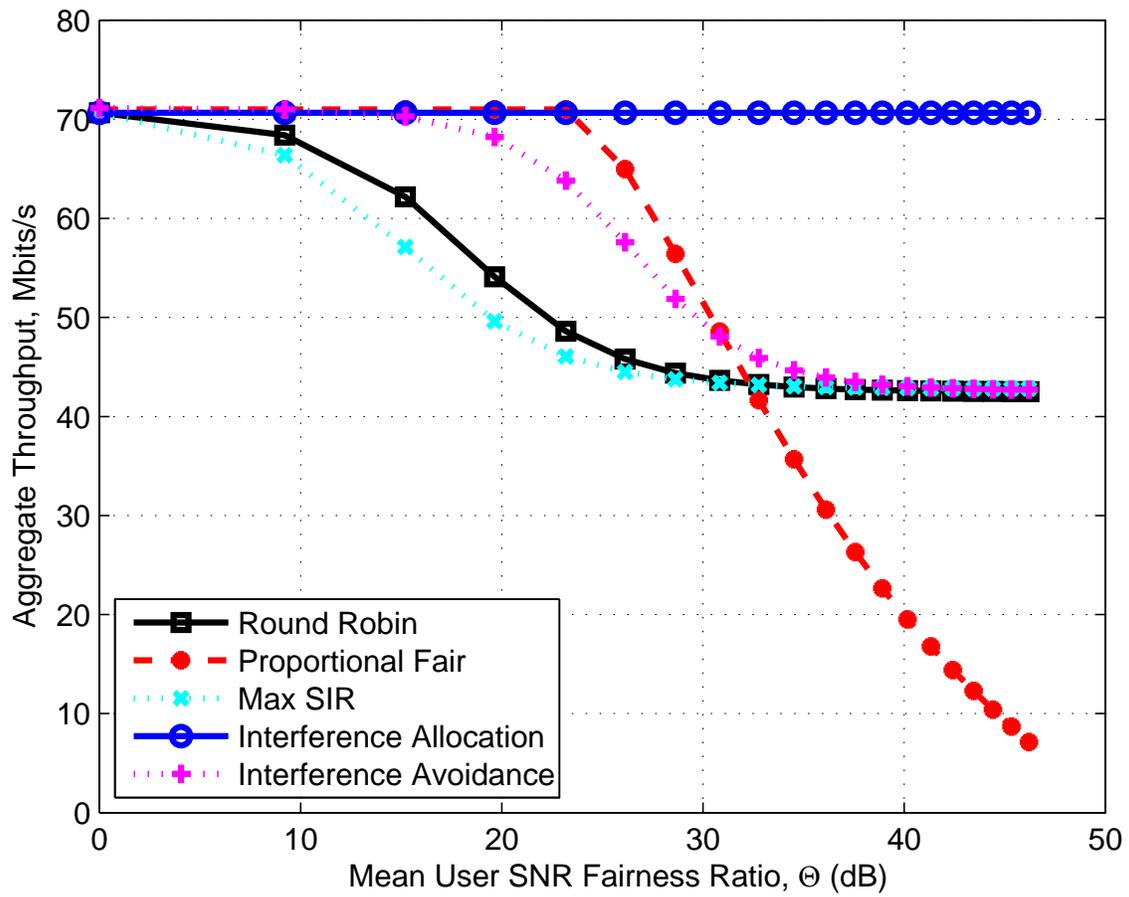

Fig. 6. Simulated aggregate BS throughput with various schedulers as a function of the unfairness ratio $\Theta$.

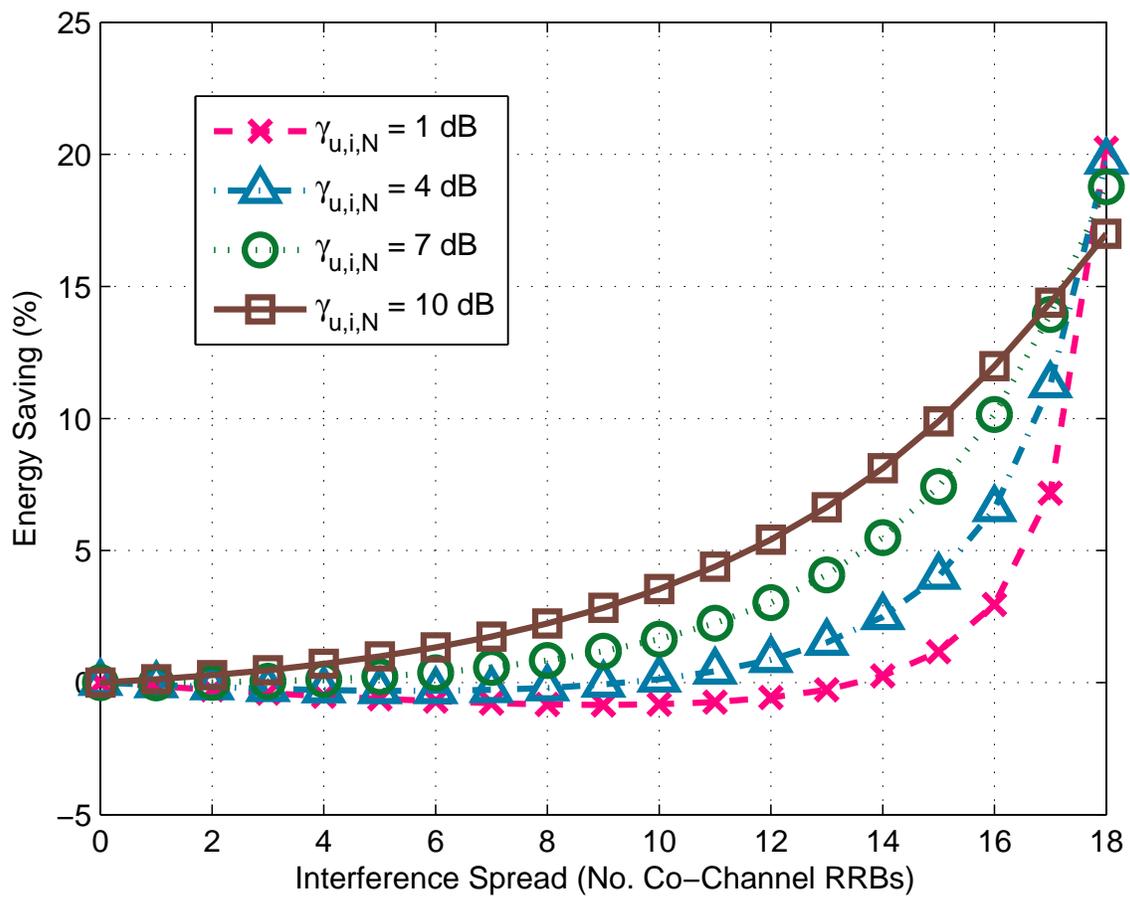

Fig. 7. Simulated energy saving compared to allocation $X$.

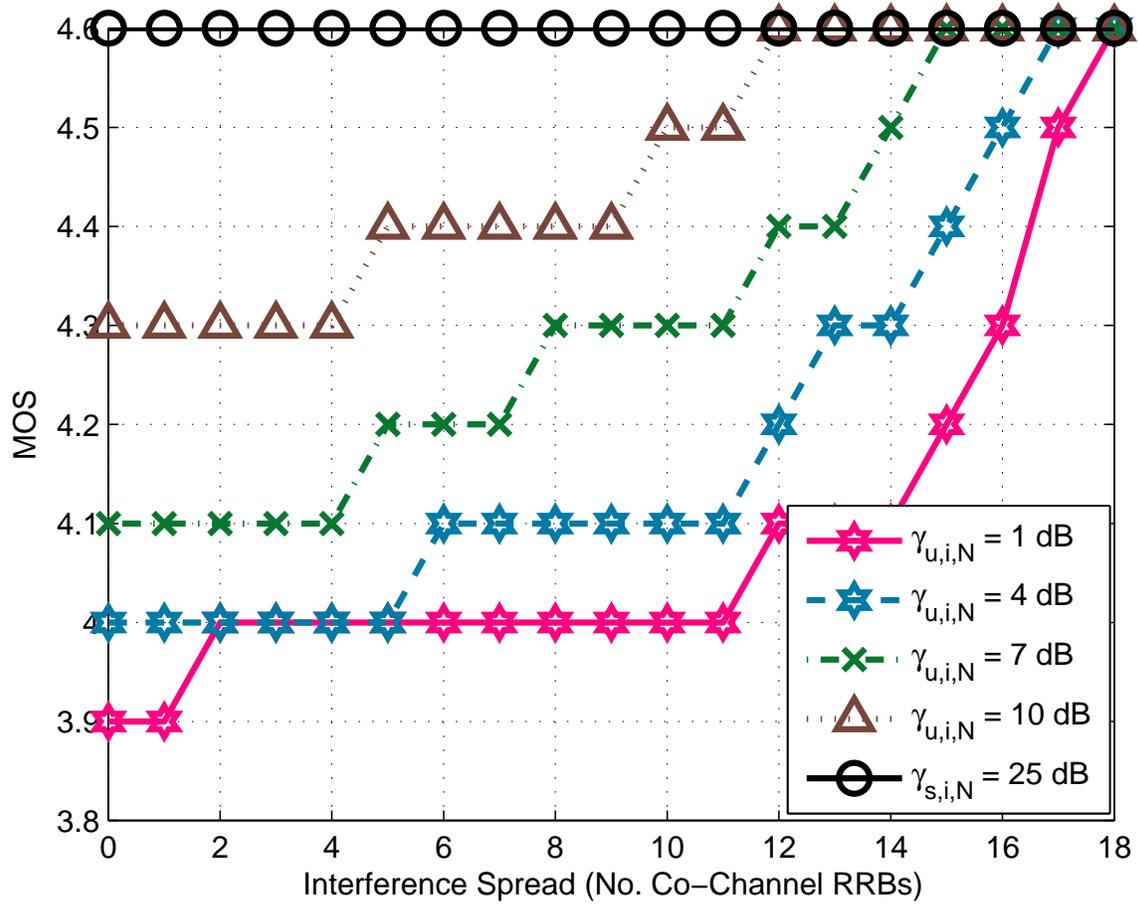

Fig. 8. Simulated UE experience (MOS) compared to allocation $X$.

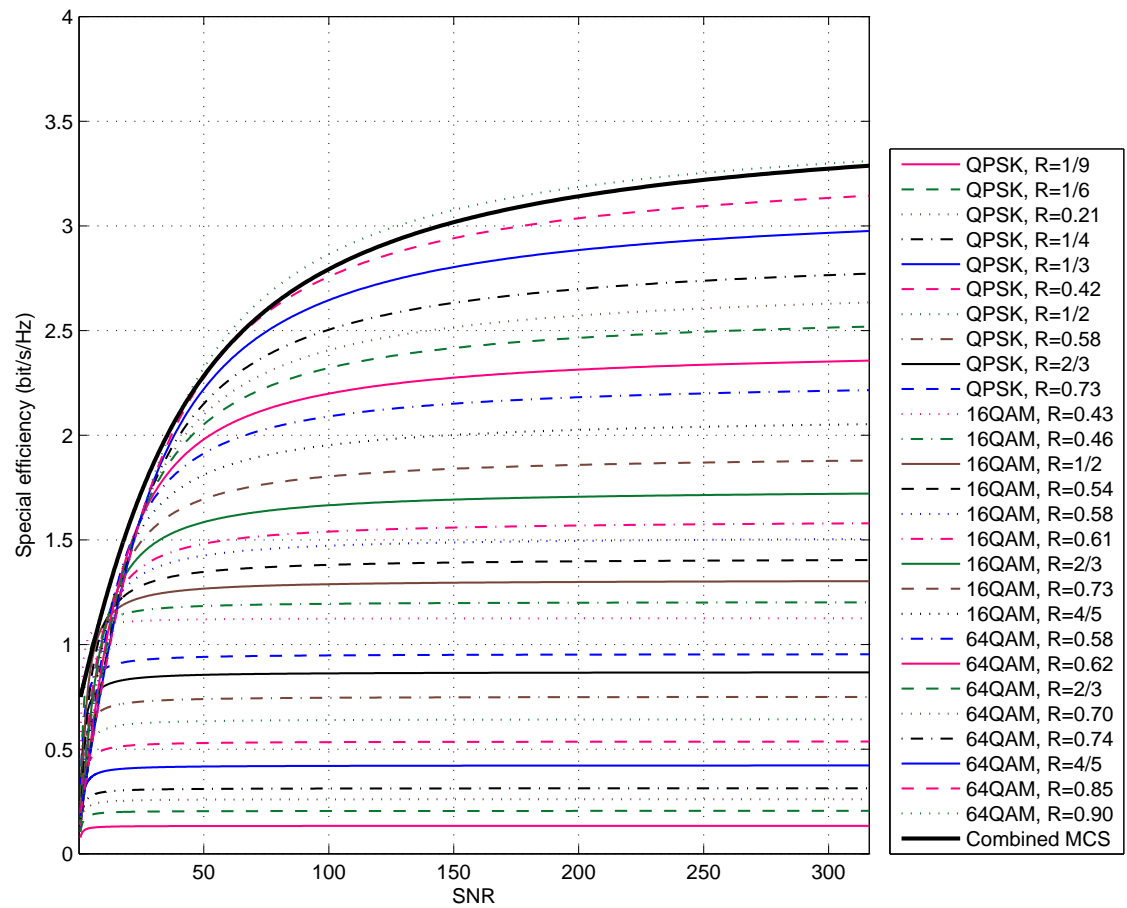

Fig. 9. Theoretical adaptive MCS capacity of LTE physical layer using expression (2), with SNR values is in linear.